\documentclass[prd,twocolumn,a4paper]{revtex4}

\usepackage{amsmath}

\begin{document}

\title{Comment on ''The Cluster Expansion for the Self-Gravitating
gas and the Thermodynamic Limit'', by de Vega and S\'anchez
(astro-ph/0307318)} 
\date{September 8, 2003}
\author{V. Laliena}
\email{laliena@posta.unizar.es}
\affiliation{Departamento de F\'{\i}sica Te\'orica, Universidad de
Zaragoza, C. Pedro Cerbuna 12, E-50009 Zaragoza (Spain)}

\maketitle

In a series of papers, de Vega and S\'anchez claimed that the thermodynamic 
limit of a self-gravitating system can be taken by letting the number of
particles, $N$, and the volume, $V$, tend to infinity keeping the ratio
$N/V^{1/3}$
constant \cite{devega}. This limit, which I call \textit{diluted} following
the terminology of the first paper of \cite{devega},
is different from the usual thermodynamic
limit, where the density $N/V$ is kept constant. The relevant variable for
the diluted limit,
which can be found by naive dimensional analysis, is $\eta=G m^2 N/V^{1/3} T$.

Recently, I proved rigorously that the diluted limit does not give a well 
defined
thermodynamic limit \cite{yo}: the relevant thermodynamic potentials are
not extensive and the thermodynamic quantities suffer from the same problems
as in the usual thermodynamic limit. For instance,
the free energy scales with $N^{5/3}$. However, in ''The Cluster Expansion
for the Self-Gravitating gas and the Thermodynamic Limit'' 
\cite{devega2}, de Vega and S\'anchez continue arguing that the diluted
limit gives extensive thermodynamic potentials and well behaved
thermodynamic quantities at sufficiently high
temperature, i.e. for $\eta>\eta_c$, where $\eta_c$ depends on the
thermodynamic ensemble as well as on the geometry of the system boundary.
To mantain these statements, these authors try to show that ''the
statements made in ref.[7] [reference \cite{yo} of the present paper]
have crucial failures which invalidate the conclusions given in ref.[7]''.
However, the argument they give is based on a misunderstunding of the 
proof of nonexistence of the diluted limit given in \cite{yo}, and
is easily refuted, as will be seen in the following.

Let us briefly remenber the proof of nonexistence of the diluted limit
given in \cite{yo}. Consider a system of $N$ classical particles, enclosed
on a region of linear size $R$ (so that $V=R^3$) interacting
via a gravitational potential conveniently regularized at short distances.
In the diluted limit we have $N\rightarrow\infty$, $R\rightarrow\infty$,
with $N/R$ constant. The variable $\eta$ of ref. \cite{devega2} is, by
definition, 
\begin{equation}
\eta=\frac{G m^2 N}{R T}\, ,
\end{equation}
since $V=R^3$ is the volume of the region available for the system.
Now, let us consider a region of linear size $R_0$, with $N\sim R_0^3$,
enclosed in the available space of the system. Note that $R_0\ll R$.
Using a simple sequence of inequalities, it is proven in \cite{yo} that
\begin{equation}
\mathcal{Z}_\mathrm{C}\geq \frac{R_0^{3N}}{N!}\,
\exp[\beta N(N-1) \kappa/R_0]\, ,
\label{ineq}
\end{equation}
where $\mathcal{Z}_\mathrm{C}$ is the canonical partition function,
$\beta=1/T$ is the inverse of the temperature, and $\kappa>0$ is 
$G m^2$ times a geometrical number independent of $R_0$ if $R_0$ is large.
The above inequality shows that the free energy grows 
with $N$ at least as $N^{5/3}$, and therefore is not extensive.

The authors of \cite{devega2} argue that since to derive  
inequality~(\ref{ineq}) I have introduced another length $R_0$,
the relevant variable is $\eta=G m^2 N/T R_0$, and, since
$R_0\sim N^{1/3}$, we have $\eta\gg \eta_c$, so that the system is deep
in the
collapse phase, where the results of \cite{devega,devega2} do not apply.

Obviously, this is wrong. $R_0$ is not a characteristic length of
the system. It is an auxiliary mathematical length, introduced just to 
prove that the diluted limit is ill-behaved. It has no physical meaning, 
and hence it is left unspecified. It can have
any value, the only retriction being that it must scale with the number
of particles as $R_0\sim N^{1/3}$. By definition, the length entering
$\eta$ in equation~(1) of ref.~\cite{devega2} is the linear size of
the spatial region available to the system. Hence, $\eta=G m^2/T R$.
Only in such case the integrals over the coordinates that appear
in equations (2.9), (2.15), etc., of \cite{devega2} can be taken
between 0 and 1. Inequality~(\ref{ineq}) can be written as
\begin{equation}
\mathcal{Z}_\mathrm{C}\geq \frac{R_0^{3N}}{N!}\,
\exp[\eta (N-1) \bar\kappa R/R_0]\, ,
\label{ineq2}
\end{equation}
where $\bar\kappa>0$ is now a dimensionless purely geometrical number 
independent of the size of the system. Since $R/R_0\sim N^{2/3}$, we 
see that the free energy scales $N^{5/3}$ if $\eta$ is kept constant,
and therefore is nonextensive.

It is clear that the grand canonical partition function cannot
scale as $\exp[N g(\eta,\mu)]$ if the canonical partition function
grows with $N$ faster than $\exp[N f(\eta)]$. Hence, the grand canonical 
ensemble cannot give an extensive thermodynamic potential, either.

Inequalities~(\ref{ineq}) or~(\ref{ineq2}) are valid whatever the value
of the temperature (i.e., of $\eta$). Hence, they prove that the gas
phase obtained in refs. \cite{devega,devega2} does not exist in the 
diluted limit. For a finite system, we expect a gas phase at high
temperature (small $\eta$) and a collapse phase at low temperature
(large $\eta$), with the two phases separated by a phase transition
or crossover at some $\eta_c$. As the system size grows, $\eta_c$
will decrease towards zero, the gas phase will shrink and the
collapse phase will eventually cover the whole phase diagram.

In \cite{devega2} it is shown that the diluted limit exists
order by order in the cluster expansion (a similar proof was
given in \cite{yo} for the high temperature expansion). 
This contradicts the proof of nonexistence of the diluted limit.
Since inequalities~(\ref{ineq}) or~(\ref{ineq2})
are derived rigorously, with no assumption, the high temperature
and cluster expansions cannot be valid. The reasons why this kind of 
expansions fail have been analyzed in \cite{yo}. Basically, there are
two possibilities:

i) The series in $\eta/N$ may not converge in the $N\rightarrow\infty$
limit, due to the contribution of high order diagrams that
are naively suppressed by powers of $1/N$, but which actually may give a 
significant contribution ought to the
short distance divergences (the cut-off behaves as $a=A/N$, where $A$
is a fixed length, \textit{Cf}. Eq. (2.8) of the paper).
Concerning this point, the statement 
that appears at the end of the introduction of the paper \cite{devega2},
''one can take the limit $N\rightarrow\infty$ and \textbf{then}
$a\rightarrow 0$'' does probably not hold due to the singularities of the
integrals that give the coefficients of the cluster expansion. The 
rigorous limit is, obviously $N\rightarrow\infty$, $a\rightarrow 0$,
with $N a$ fixed. The modification of the procedure to take the
$N\rightarrow\infty$ limit may be at the core of the failure of the 
cluster expansion developed in \cite{devega2}.

ii) Even if the
cluster expansion were convergent, the series could not represent the
thermodynamic potential, since it is in contradiction with the
rigorous result of \cite{yo}. In deriving the cluster expansion
there is at least one mathematically unjustified exchange of limits
that may invalidate the equality between the thermodynamic potential
and the cluster series. The cluster expansion (I use the notation of
\cite{devega2}),
\begin{equation}
Q_N(\eta)=1+\sum \int f_{ij}+\sum \int f_{ij} f_{kl}+\ldots\, ,
\end{equation}
is rigorous, since the number of terms in the sums is finite for finite
$N$. To proceed further, the authors of \cite{devega2} expand
$f_{ij}$ in powers of $\eta/N$, and exchange the sum and integral.
Mathematically, it can be very difficult to analyze the conditions
under which this exchange of limits is allowed, but we can get some
insight from physical intuition. Inequality~(\ref{ineq2}) suggests that
the system is collapsed for large $N$ if $\eta$ is of order 1 respect to
$1/N$. This means that the cluster expansion is dominated by the
higher order terms (many particles within a cluster). Hence,
the canonical (Gibbs) integration measure on configuration space
is very concentrated on collapsed configurations. Hence, this measure
is very different from the flat measure $\prod_i d^3 r_i$ that
correspond to a gas. Keeping only the dominant 
of the expansion of $f_{ij}$ in powers of $\eta/N$
means that one is using effectively the flat measure. To recover 
something similar to the concentrated true measure, one has to sum an 
infinite number of terms in $\eta/N$. In other words, the expansion
in $\eta/N$ is valid only in the gas phase. Inequality~(\ref{ineq2}) 
suggests that the gas phase can only take place for $\eta$ of the order of
$N^{-2/3}$. Hence, the radius of convergence of the expansion
in $\eta$ shrinks to zero as $N\rightarrow\infty$.

Similar statements claiming the existence of the diluted
thermodynamic limit have been made in \cite{devega3} without even 
mentioning the results of ref.~\cite{yo}.

\vfill\eject 

\end{document}